\documentclass[dvipdfmx]{pasj01}
\usepackage{multirow}

\begin{document}
\Received{2015/06/19}
\Accepted{2015/10/14}

\title{New Detections of Galactic Molecular Absorption Systems toward ALMA Calibrator Sources}

\author{Ryo \textsc{Ando},\altaffilmark{1}
        Kotaro \textsc{Kohno},\altaffilmark{1}
        Yoichi \textsc{Tamura},\altaffilmark{1}
        Takuma \textsc{Izumi},\altaffilmark{1}
        Hideki \textsc{Umehata},\altaffilmark{1,2}
        and
        Hiroshi \textsc{Nagai}\altaffilmark{3}
    }
        
\altaffiltext{1}{Institute of Astronomy, The University of Tokyo, 2-21-1 Osawa, Mitaka, Tokyo 181-0015, Japan }
\email{ando@ioa.s.u-tokyo.ac.jp}
\altaffiltext{2}{European Southern Observatory, Karl Schwarzschild Str. 2, D-85748 Garching, Germany}
\altaffiltext{3}{National Astronomical Observatory of Japan, 2-21-1 Osawa, Mitaka, Tokyo 181-8588, Japan }

\KeyWords{quasars: absorption lines --- ISM: molecules --- ISM: abundances --- radio lines: ISM} 

\maketitle

\begin{abstract}
We report on Atacama Large Millimeter/submillimeter Array (ALMA) detections 
of molecular absorption lines in Bands 3, 6 and 7 toward four radio-loud quasars,
which were observed as the bandpass and complex gain calibrators. 
The absorption systems, three of which are newly detected, 
are found to be Galactic origin.
Moreover, HCO absorption lines toward two objects are detected, 
which almost doubles the number of HCO absorption samples in the Galactic diffuse medium.
In addition, high HCO to H$^{13}$CO$^+$ column density ratios are found, 
suggesting that the interstellar media (ISM) observed toward the two calibrators 
are in photodissociation regions, 
which observationally illustrates the chemistry of diffuse ISM driven by ultraviolet (UV) radiation.
These results demonstrate that calibrators in the ALMA Archive are potential sources 
for the quest for new absorption systems and for detailed investigation of the nature of the ISM.
\end{abstract}

\section{Introduction}
A molecular absorption system, in which molecular absorption occurs
along the line-of-sight toward a radio-loud quasar, 
is a powerful tool for the investigation of the chemistry and physics of interstellar media (ISM),  
from those within Galactic diffuse clouds to high-redshift ($z$) galaxies.
Investigations of the ISM at low excitation states in a distant system 
indicate higher sensitivity toward absorption than emission 
(\cite{WC94}; \cite{WC96}; \cite{Muller+11}), 
as the absorption line depths are independent of the distance to the system, 
but dependent on the brightness of the background sources (\cite{Muller+14}).

\begin{table*}[t]
  \tbl{Summary of ALMA archival data in which molecular absorption systems have been found.
  The objects toward which molecular absorption lines were detected are indicated in bold.}{
    \begin{tabular}{ccccccc}
   ~~ \\
\hline\noalign{\vskip3pt} 
      \multirow{2}{*}{Project} & Observation & \multirow{2}{*}{Band}
       & Bandpass & Integration & Complex gain & Integration \\
       & date & & calibrator & time (s) & calibrator & time (s) \\ [2pt] \hline\noalign{\vskip3pt}
       2011.0.00217.S & July 29 - August 1 2012 & 3 & \bf{NRAO530} 
       & 131 & \bf{J1717$-$337} & 107 \\
       2011.0.00351.S & April 7 - July 31 2012 & 3 & 
       J1924$-$292/3C279 & 421/104 & \bf{J1717$-$337} & 333/88 \\
       2011.0.00259.S & May 8 2012 & 3 & 3C279 & 21 & \bf{J1625$-$254} & 55 \\
       2011.0.00531.S & March 27 - May 4 2012 & 6 & 
       J1924$-$292/3C279 & 133/132 & \bf{J1625$-$254} & 61/61 \\
       2011.0.00733.S & August 1 2012 & 3 & 
       {\bf NRAO530} & 85 & {\bf J1604$-$446} & 69 \\
       & June 18 - July 4 2012 & 6 & {\bf NRAO530} & 87 & {\bf J1604$-$446} & 138 \\
       2011.0.00524.S & January 12 2012 & 7 & J1427$-$421 & 103 & \bf{J1604$-$446} & 239 \\ [2pt] 
      \hline
    \end{tabular}}\label{tab1}
\end{table*}

\begin{table*}[t]
\tbl{List of Galactic molecular absorption systems detected in this work. 
The objects and detected species indicated in bold are systems and molecules newly detected, 
respectively. The uncertainty of the absolute flux scales is typically $\sim 10$\%.}{
\begin{tabular}{ccccccccc}
   ~~ \\  
\hline\noalign{\vskip3pt} 
\multicolumn{1}{c}{Object}
      & \multicolumn{2}{c}{Galactic coordinates} 
      & \multirow{2}{*}{Band}
      & \multicolumn{1}{c}{Velocity resolution}
      & \multicolumn{1}{c}{RMS} 
      & \multicolumn{1}{c}{Continuum} 
      & \multicolumn{1}{c}{Detected species in this work}\\ 
name & $ l \ (^\circ)$ & $b \ (^\circ)$  &  & $\Delta v$ (km s$^{-1}$) 
& $\sigma$ (mJy) & flux (Jy) & ({\bf Bold}: new-detection) \\ [2pt] \hline\noalign{\vskip3pt} 
 & & & & & & & {\bf c-C$_3$H$_2$(2$_{(1,2)}$--1$_{(0,1)}$), HCS$^+$(2--1), H$^{13}$CN(1--0), }\\
{\bf J1717$-$337} & \ 352.73 \ & 2.39 & 3 &
3.4 & 3 & 1.4 & {\bf HCO(1$_{(0,1)}$--0$_{(0,0)}$), H$^{13}$CO$^+$(1--0), HN$^{13}$C(1--0), } \\
 & & & & & & & {\bf C$_2$H(1--0), HCN(1--0), HCO$^+$(1--0), CS(2--1)} \\ [2pt] \hline\noalign{\vskip3pt}
\multirow{2}{*}{{\bf J1625$-$254}} & \multirow{2}{*}{352.14} & \multirow{2}{*}{16.32} & 3 & 3.4 & 7 
 & 1.3 & {\bf c-C$_3$H$_2$(2$_{(1,2)}$--1$_{(0,1)}$), C$_2$H(1--0), HCN(1--0)} \\
 & & & 6 & 1.3 & 19 & 0.69 & {\bf CO(2--1)} \\ [2pt] \hline\noalign{\vskip3pt}
 & & & 3 & 3.4 & 6 & 0.75 & {\bf CS(2--1)} \\
{\bf J1604$-$446} & 335.16 & 5.76 & 6 & 1.3 & 5 & 0.39 & {\bf CO(2--1)} \\
 & & & 7 & 0.86 & 17 & 0.60 & {\bf CO(3--2)} \\ [2pt] \hline\noalign{\vskip3pt}
 & & &  \multirow{2}{*}{3} & \multirow{2}{*}{3.4} & \multirow{2}{*}{1}
 & \multirow{2}{*}{0.94} & {\bf HCO(1$_{(0,1)}$--0$_{(0,0)}$)}, H$^{13}$CO$^+$(1--0), SiO(2--1), \\ 
NRAO530 & 12.03 & 10.81 & & & & & C$_2$H(1--0), HCN(1--0), HCO$^+$(1--0) \\
 & & & 6 & 1.3 & 5 & 0.77 & CO(2--1) \\  [2pt]  
\hline
\end{tabular}}\label{tab2}
\end{table*}

Molecular absorption line systems at high-$z$ play an important role 
in the detection of molecules at high-$z$ with high sensitivity, 
and in investigation of their chemical evolution 
(\cite{Combes08} (a review); \cite{WC96}; \cite{Muller+14}).
In addition, observations of such distant systems can contribute to fundamental physics knowledge
by allowing the cosmological variation of fundamental constants or 
the cosmic microwave background (CMB) temperature as a function of $z$ to be probed
(\cite{Combes08}; \cite{Henkel+09}; \cite{Muller+11}).
However, note that only four objects are known to have high-$z$ molecular absorption 
at millimeter wavelength (\cite{Combes08}; \cite{Curran+11}).
Observation of molecular absorption lines in quasar host galaxies 
also facilitates investigation of gas structures around the active galactic nuclei (AGNs) 
of edge-on host galaxies (\cite{Roberts70}; \cite{Israel+91}; \cite{Espada+10}).
Furthermore, absorption studies of such host galaxies may be useful 
in facilitating the direct observation of the putative molecular tori in AGNs, 
although no detection of this nature has been successfully performed to date 
(\cite{Okuda+13} and references therein).

Even the molecular absorption inside our Galaxy is of considerable interest.
Observations of molecular absorption lines in diffuse ISM 
can improve our understanding of its chemistry.
For example, such observations can allow determination of the isotope ratios 
of fundamental elements based on isotopologue column density ratios (\cite{LL98}), 
and may reveal the chemical richness of diffuse ISM by extending the molecular inventory 
(e.g., \cite{Liszt+14}).
It is important to investigate the physical and chemical states of diffuse ISM, 
because such extended diffuse media are thought to be dominant 
in the interarm regions of our Galaxy (\cite{Sawada+12}), 
and even account for a certain proportion of the total gas in our Galaxy (\cite{Pineda+13}).
High spatial-resolution observation of the nearby spiral galaxy, M51, also suggests 
that a diffuse thick disk of molecular gas extends broadly over that galaxy (\cite{Pety+13}).
Therefore, investigation of the properties of diffuse ISM 
may even facilitate further comprehension of galactic evolution.
However, the detailed nature of diffuse molecular gas is poorly understood, 
primarily because emission studies experience critical difficulties 
in detecting such diffuse gases
because of their low densities and low excitation temperatures (\cite{Lequeux+93}).
Several diffuse ISM heating mechanisms have been advocated in previous studies, 
for example, those based on turbulence (\cite{Godard+10}) 
and UV radiation from OB-stars (\cite{HT99}); 
however, there exist a limited number of studies observationally 
supporting such mechanisms 
from the point of view of the chemical states of diffuse molecular gases.
Meanwhile, absorption observations can potentially be used to examine the properties of diffuse ISM.

\begin{table*}[t]
  \tbl{Absorption line decomposition products of newly detected species.}{
    \begin{tabular}{cccrccc}
   ~~ \\
\hline\noalign{\vskip3pt} 
      Object & \multirow{2}{*}{Species} & Frequency & LSR velocity \ \ 
      & Peak optical depth \footnotemark[$*1$] & FWHM & Column density \footnotemark[$*1$] \\
      name & & $\nu$ (GHz) & $V_{\mathrm{LSR}}$ (km s$^{-1}$) 
      & $\tau_{\nu}$ & $\Delta V$ (km s$^{-1}$) 
      & $N_{\mathrm{total}}$ (cm$^{-2}$) \\ [2pt] \hline\noalign{\vskip3pt}
      & c-C$_3$H$_2$(2$_{(1,2)}$--1$_{(0,1)}$) & 85.3389 & $-2.82 \pm 0.18$ & $0.054 \pm 0.004$ 
      & $5.62 \pm 0.44$ & $(1.5 \pm 0.2) \times 10^{12}$ \\
      & & & $6.81 \pm 0.50$ & $0.017 \pm 0.003$ 
      & $5.66 \pm 1.34$ & $(4.8 \pm 1.4) \times 10^{11}$ \\
      & & & $-20.73 \pm 0.17$ & $0.012 \pm 0.002$ 
      & $6.12 \pm 0.43$ & $(3.7 \pm 0.7) \times 10^{11}$ \\
      & HCS$^+$(2--1) & 85.3479 & $-2.26 \pm 0.53$ & $0.006 \pm 0.002$ 
      & $5.42 \pm 1.25$ & $(3.0 \pm 1.2) \times 10^{11}$ \\
      & H$^{13}$CN(1--0) \footnotemark[$*2$]  & 86.3402 & $-3.42 \pm 0.22$ & $0.013 \pm 0.001$ 
      & $5.56 \pm 0.59$ & $(1.6 \pm 0.2) \times 10^{11}$ \\
      & HCO(1$_{(0,1)}$-0$_{(0,0)}$) \footnotemark[$*2$] 
      & 86.6708 & $-3.03 \pm 0.10$ & $0.016 \pm 0.002$ 
      & $5.33 \pm 0.23$ & $(2.1 \pm 0.2) \times 10^{12}$ \\ 
      & H$^{13}$CO$^+$(1--0) & 86.7543
      & $-2.64 \pm 0.08$ & $0.011 \pm 0.002$ 
      & $6.33 \pm 0.21$ & $(8.6 \pm 1.2) \times 10^{10}$ \\
      & & & $5.78 \pm 0.80$ & $0.005 \pm 0.002$ 
      & $3.76 \pm 2.16$ & $(2.3 \pm 1.6) \times 10^{10}$ \\
      & HN$^{13}$C(1--0) & 87.0909 & $-3.57 \pm 0.14$ & $0.006 \pm 0.001$ 
      & $5.25 \pm 0.33$ & $(6.4 \pm 1.4) \times 10^{10}$ \\
      J1717$-$337 & C$_2$H(1--0) \footnotemark[$*2$] & 87.3169
      & $-2.60 \pm 0.09$ & $0.097 \pm 0.004$ 
      & $5.55 \pm 0.22$ & $(3.7 \pm 0.2) \times 10^{13}$ \\
      & & & $7.10 \pm 0.33$ & $0.035 \pm 0.005$ 
      & $4.56 \pm 0.81$ & $(1.1 \pm 0.3) \times 10^{13}$ \\
      & & & $-20.76 \pm 0.07$ & $0.024 \pm 0.001$ 
      & $4.39 \pm 0.17$ & $(7.1 \pm 0.5) \times 10^{12}$ \\
      & & & $-11.53 \pm 0.53$ & $0.007 \pm 0.001$ 
      & $5.55 \pm 1.44$ & $(2.8 \pm 0.9) \times 10^{12}$ \\
      & HCN(1--0) \footnotemark[$*2$] & 88.6318 & $-3.22 \pm 0.25$ & $0.628 \pm 0.064$ 
      & $6.52 \pm 0.75$ & $(8.3 \pm 1.3) \times 10^{12}$ \\
      & & & $-20.84 \pm 0.20$ & $0.022 \pm 0.002$ 
      & $5.63 \pm 0.51$ & $(2.5 \pm 0.3) \times 10^{11}$ \\
      & HCO$^+$(1--0) & 89.1885 & $-2.54 \pm 0.17$ & $0.769 \pm 0.059$ 
      & $6.91 \pm 0.40$ & $(6.3 \pm 0.6) \times 10^{12}$ \\
      & & & $6.93 \pm 0.11$ & $0.206 \pm 0.011$ 
      & $4.42 \pm 0.29$ & $(1.1 \pm 0.1) \times 10^{12}$ \\
      & & & $-21.54 \pm 0.08$ & $0.056 \pm 0.002$ 
      & $4.84 \pm 0.19$ & $(3.2 \pm 0.2) \times 10^{11}$ \\
      & & & $-11.76 \pm 0.36$ & $0.023 \pm 0.002$ 
      & $5.77 \pm 0.95$ & $(1.5 \pm 0.3) \times 10^{11}$ \\
      & CS(2--1) & 97.9810 & $-3.29 \pm 0.04$ & $0.084 \pm 0.007$ 
      & $3.71 \pm 0.10$ & $(2.7 \pm 0.2) \times 10^{12}$ \\ [2pt] \hline\noalign{\vskip3pt} 
      \multirow{4}{*}{J1625$-$254} & c-C$_3$H$_2$(2$_{(1,2)}$--1$_{(0,1)}$) & 85.3389 
      & $6.50 \pm 0.12$ & $0.089 \pm 0.006$ 
      & $5.15 \pm 0.28$ & $(2.3 \pm 0.2) \times 10^{12}$ \\
      & C$_2$H(1--0) \footnotemark[$*2$]  & 87.3169 & $5.12 \pm 0.10$ & $0.134 \pm 0.006$ 
      & $5.36 \pm 0.24$ & $(4.9 \pm 0.3) \times 10^{13}$ \\
      & HCN(1--0) \footnotemark[$*2$]  & 88.6318 & $6.93 \pm 0.26$ & $0.097 \pm 0.037$ 
      & $6.84 \pm 0.57$ & $(1.4 \pm 0.5) \times 10^{12}$ \\
      & CO(2--1) & 230.5380 & $5.32 \pm 0.12$ & $0.143 \pm 0.032$ 
      & $2.98 \pm 0.28$ & $(1.6 \pm 0.4) \times 10^{15}$ \\ [2pt] \hline\noalign{\vskip3pt}
      \multirow{4}{*}{J1604$-$446} & CS(2--1) & 97.9810 & $-46.92 \pm 0.18$ & $0.086 \pm 0.009$ 
      & $4.32 \pm 0.41$ & $(3.2 \pm 0.5) \times 10^{12}$ \\
      & CO(2--1) & 230.5380 & $-41.54 \pm 0.03$ & $2.328 \pm 0.329$ 
      & $1.82 \pm 0.07$ & \hspace{2.75mm} $(1.1 \pm 0.2) \times 10^{16}$ \footnotemark[$*3$] \\ 
      & & & $-38.73 \pm 0.05$ & $0.467 \pm 0.022$ 
      & $2.55 \pm 0.13$ & \hspace{2.75mm} $(3.2 \pm 0.2) \times 10^{15}$ \footnotemark[$*3$] \\
      & CO(3--2) & 345.7960 & $-41.52 \pm 0.03$ & $1.034 \pm 0.117$ 
      & $1.16 \pm 0.07$ & \hspace{2.75mm} $(1.1 \pm 0.2) \times 10^{16}$ \footnotemark[$*3$] 
      \\ [2pt] \hline\noalign{\vskip3pt}
      NRAO530 & HCO(1$_{(0,1)}$--0$_{(0,0)}$) \footnotemark[$*2$] 
      & 86.6708 & $6.19 \pm 0.18$ & $0.005 \pm 0.001$ 
      & $3.41 \pm 0.40$ & $(3.7 \pm 0.7) \times 10^{11}$ \\  [2pt] 
      \hline
    \end{tabular}}\label{tab3}
    \begin{tabnote}
    \hangindent6pt\noindent
\hangindent0pt\noindent
\hbox to8pt{\footnotemark[$*1$]\hss}\unskip
The errors of the peak optical depths and column densities stem from the uncertainty involved 
in the flux calibration, RMS noise calculation, and line fits.
\newline
\hangindent0pt\noindent
\hbox to8pt{\footnotemark[$*2$]\hss}\unskip
The data for the molecular lines with hyperfine structure components consist of the strongest values.
\newline
\hangindent0pt\noindent
\hbox to8pt{\footnotemark[$*3$]\hss}\unskip
Using a rotation diagram with two transition lines, 
the total column density and $T_{\mathrm{ex}}$ of the main CO component are determined to be 
$1.1 \times 10^{16}$ cm$^{-2}$ and 6.2 K, respectively, 
and the $T_{\mathrm{ex}}$ of the weaker CO velocity component is also assumed to be 6.2 K, 
although these results should be treated carefully (see section 3).
\end{tabnote}
\end{table*}

\begin{table*}[t]
  \tbl{Upper limits on peak optical depths and column densities of undetected species.
  The $T_{\mathrm{ex}}$ of all the molecules below are assumed to be 2.73 K.}{
    \begin{tabular}{cccccc}
   ~~ \\
\hline\noalign{\vskip3pt} 
      Object & \multirow{2}{*}{Species} & Frequency 
      & Peak optical depth & Assumed FWHM & Column density ($3\sigma$ upper limit) \\
      name & & $\nu$ (GHz) & ($3\sigma$ upper limit) $\tau_{\nu}$ & $\Delta V$ (km s$^{-1}$) 
      & $N_{\mathrm{total}}$ (cm$^{-2}$) \\ [2pt] \hline\noalign{\vskip3pt}
      \multirow{5}{*}{J1717$-$337} & HC$^{18}$O$^+$(1--0) & 85.1622 & 0.006 
      & 5.62 \footnotemark[$*2$] & $4.2 \times 10^{10}$  \\
      & HC$^{15}$N(1--0) & 86.0550 & 0.006 & 5.62 \footnotemark[$*2$] & $7.1 \times 10^{10}$  \\
      & SiO(2--1) & 86.8470 & 0.005 & 5.62 \footnotemark[$*2$] & $9.3 \times 10^{10}$  \\
      & HC$^{17}$O$^+$(1--0) & 87.0575 & 0.005 & 5.62 \footnotemark[$*2$] 
      & $3.1 \times 10^{10}$  \\
      & HOC$^+$(1--0) & 89.4874 & 0.007 & 5.62 \footnotemark[$*2$] 
      & $8.7 \times 10^{10}$  \\ [2pt] \hline\noalign{\vskip3pt} 
      \multirow{8}{*}{J1625$-$254}& HCS$^+$(2--1) & 85.3479 & 0.015 
      & 5.78 \footnotemark[$*3$] & $8.2 \times 10^{11}$  \\
      & H$^{13}$CN(1--0) \footnotemark[$*1$] & 86.3402 & 0.015 & 5.78 \footnotemark[$*3$] 
      & $1.8 \times 10^{11}$  \\
      & HCO(1$_{(0,1)}$-0$_{(0,0)}$) \footnotemark[$*1$] & 86.6708 & 0.015 
      & 5.78 \footnotemark[$*3$] & $2.1 \times 10^{12}$  \\ 
      & H$^{13}$CO$^+$(1--0) & 86.7543 & 0.015 & 5.78 \footnotemark[$*3$] 
      & $1.1 \times 10^{11}$  \\ 
      & SiO(2--1) & 86.8470 & 0.015 & 5.78 \footnotemark[$*3$] & $3.1 \times 10^{11}$ \\
      & CS(2--1) & 97.9810 & 0.017 & 5.78 \footnotemark[$*3$] & $8.4 \times 10^{11}$ \\
      & C$^{18}$O(2--1) & 219.5604 & 0.163 & 2.98 \footnotemark[$*4$] & $1.9 \times 10^{15}$ \\
      & $^{13}$CO(2--1) & 220.3987 & 0.196 & 2.98 \footnotemark[$*4$] & $2.3 \times 10^{15}$ 
      \\ [2pt] \hline\noalign{\vskip3pt}
      J1604$-$446 & C$^{34}$S(2--1) & 96.4130 & 0.025 
      & 4.32 \footnotemark[$*5$] & $9.3 \times 10^{11}$ \\ [2pt] 
      \hline
    \end{tabular}}\label{tab4}
    \begin{tabnote}
    \hangindent6pt\noindent
\hangindent0pt\noindent
\hbox to8pt{\footnotemark[$*1$]\hss}\unskip
The data for the molecular lines with hyperfine structure components consist of the strongest values.
\newline
\hangindent0pt\noindent
\hbox to8pt{\footnotemark[$*2$]\hss}\unskip
The average of FWHM value of all the detected molecular lines toward J1717$-$337.
\newline
\hangindent0pt\noindent
\hbox to8pt{\footnotemark[$*3$]\hss}\unskip
The average of FWHM value of all the detected molecular lines toward J1625$-$254 in Band 3.
\newline
\hangindent0pt\noindent
\hbox to8pt{\footnotemark[$*4$]\hss}\unskip
Assuming the same FWHM as the CO(2--1) line toward J1625$-$254.
\newline
\hangindent0pt\noindent
\hbox to8pt{\footnotemark[$*5$]\hss}\unskip
Assuming the same FWHM as the CS(2--1) line toward J1604$-$446.
\end{tabnote}
\end{table*}

However, the number of known molecular absorption systems is rather limited.
In regards to Galactic molecular absorption at millimeter wavelength,
$\sim 30$ systems have been noted (\cite{LL96}; \cite{LL00}), 
whereas the lines-of-sight toward a small number of bright sources, 
such as 3C111, BL Lac, and NRAO530, have been studied extensively 
(e.g., \cite{LL98}; \cite{Liszt+14}).
In addition, previous searches for molecular absorption systems have been limited
in terms of sensitivity and velocity resolution.
Some of these investigations have reported non-detection (e.g. \cite{Curran+11}), 
even though such studies have been useful in detecting rare molecular species 
(\cite{Friedel+11}).

The advent of the Atacama Large Millimeter/submillimeter Array (ALMA)
offers high sensitivity and velocity resolution, 
allowing researchers to obtain high-quality spectra even within short exposure times.
The ALMA archival data offer a vast number of sensitive quasar spectra 
that have been primarily obtained for bandpass and complex gain calibration.
Thus, it is possible that ALMA has detected molecular absorption lines 
within the spectra of its calibrator sources, 
and such cases would equate to discoveries of new systems
\textit{without} any additional observation.

In this paper, we present the first results of detections of molecular absorption systems 
using ALMA calibrator sources.
In section 2, we show the data analysis method and,
in section 3, we report on the detection of four molecular absorption systems within the Galaxy.
In section 4, we discuss the HCO absorption lines, 
as HCO is the most noteworthy molecule detected in this study. 

\section{Data Analysis}
We select 36 calibrator sources from the whole sky, which fulfill the following criteria:
(a) The data are available in the ALMA archive from prior to late 2014, i.e., Cycle 0 data;
(b) The continuum flux is detected at $> 0.2$ Jy,
so a high signal-to-noise ratio (S/N) can be expected in Bands 3, 6, or 7; and
(c) The data are taken at a frequency resolution of $< 1$ MHz in frequency division mode (FDM).
Information on all the sources analyzed in the study is given in Appendix 1.
We create three-dimensional (3D) cubes from the calibrated visibilities 
of the bandpass and complex gain calibrators 
using the Common Astronomy Software Applications (CASA) package \verb|CLEAN| task,
with Briggs weighting (with a \verb|robust| parameter of 0.5).
Although the spectra of the majority of the calibrators exhibit featureless continua, 
significant absorption lines ($> 3\sigma$) are detected
in four objects observed in the six projects listed in table \ref{tab1}. 
Detailed information on these objects is given in table \ref{tab2}.

When a bandpass calibrator also exhibits absorption lines, 
the calibration process is performed once more.
The bandpass-calibrator frequency channels showing absorption lines
(typical widths of $< 5$ km s$^{-1}$ or $< 2.5$ MHz) are flagged 
and linearly interpolated using nearby channels,
as interpolation of this nature with a width of $\lesssim 20$ MHz
does not seriously affect the calibration processes.
Then, the revised bandpass calibration is applied 
so as to yield the final spectrum of the complex gain calibrator.
The spectrum of the flux calibrator (the Neptune in this case) 
to which the interpolated bandpass calibration is applied 
does not exhibit any artificial features at the frequencies where the lines are found, 
suggesting that our bandpass interpolation is valid.

The continuum spectrum is subtracted from that of the detected absorption lines
using the \verb|IMCONTSUB| task,
and the line spectrum is then divided by the continuum spectrum using the \verb|IMMATH| task
(we have confirmed that the same continuum subtraction is conducted
in both the image and $uv$ planes).
Therefore, the line/continuum flux density ratio values are in the $-1$ - 0 range. 
The normalized line spectrum is then Gaussian fitted 
to allow measurement of the central velocity, full width at half maximum (FWHM), 
and the absorption line depth, which are listed in table \ref{tab3}.
The velocity components are identified 
by comparing the profiles of the hyperfine structure lines and/or the multi-transition lines.

\begin{figure*}[p]
 \begin{center}
  \includegraphics[width=160mm]{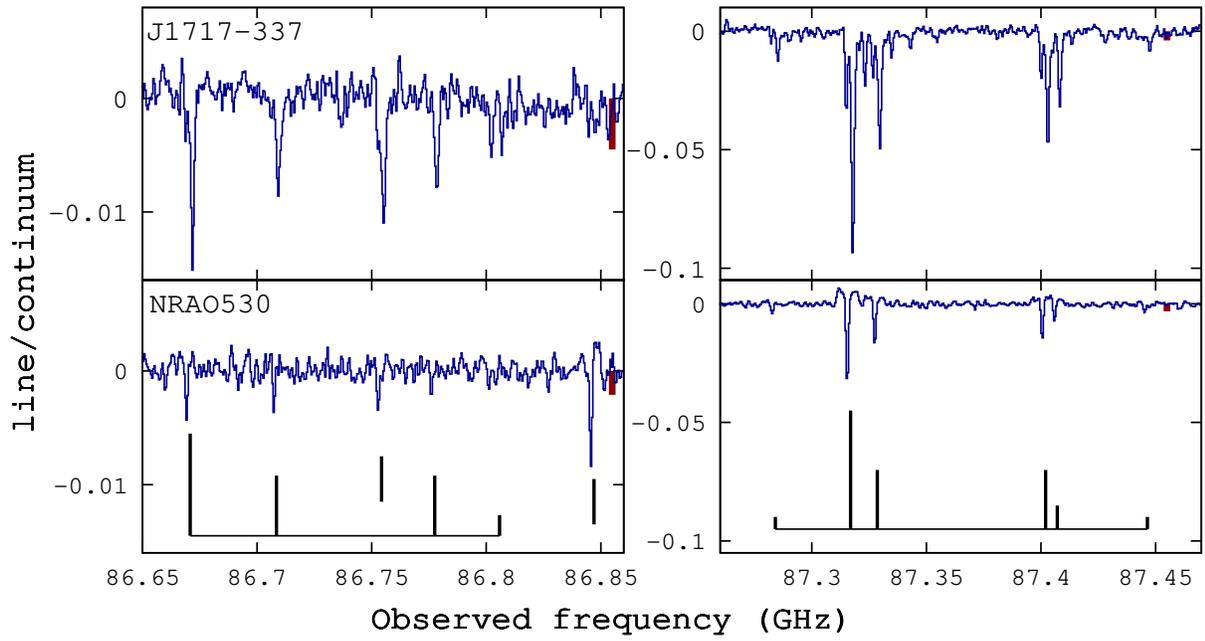} 
 \end{center}
\caption{J1717$-$337 and NRAO530 spectra, 
  shown in the top and bottom rows, respectively.
  The $3\sigma$ noise level is indicated in a red bar at the top-right of each box 
  (this approach is used hereinafter).
  The line rest frequencies and intrinsic strength ratios of the hyperfine components of 
  HCO(1$_{(0,1)}$--0$_{(0,0)}$) and C$_2$H(1--0) are shown in the bottom row, from left to right. 
  In the bottom-left box, the line rest frequencies of H$^{13}$CO$^+$(1--0) (86.7543 GHz) 
  and SiO(2--1) (86.8470 GHz) are also indicated in bars.
  Although the lines in the NRAO530 spectra seem to have weak emissions, 
  no sign of emission is confirmed around the source in its 3D data cubes.
  }\label{fig1}
\end{figure*}

\begin{figure*}[p]
 \begin{center}
  \includegraphics[width=160mm]{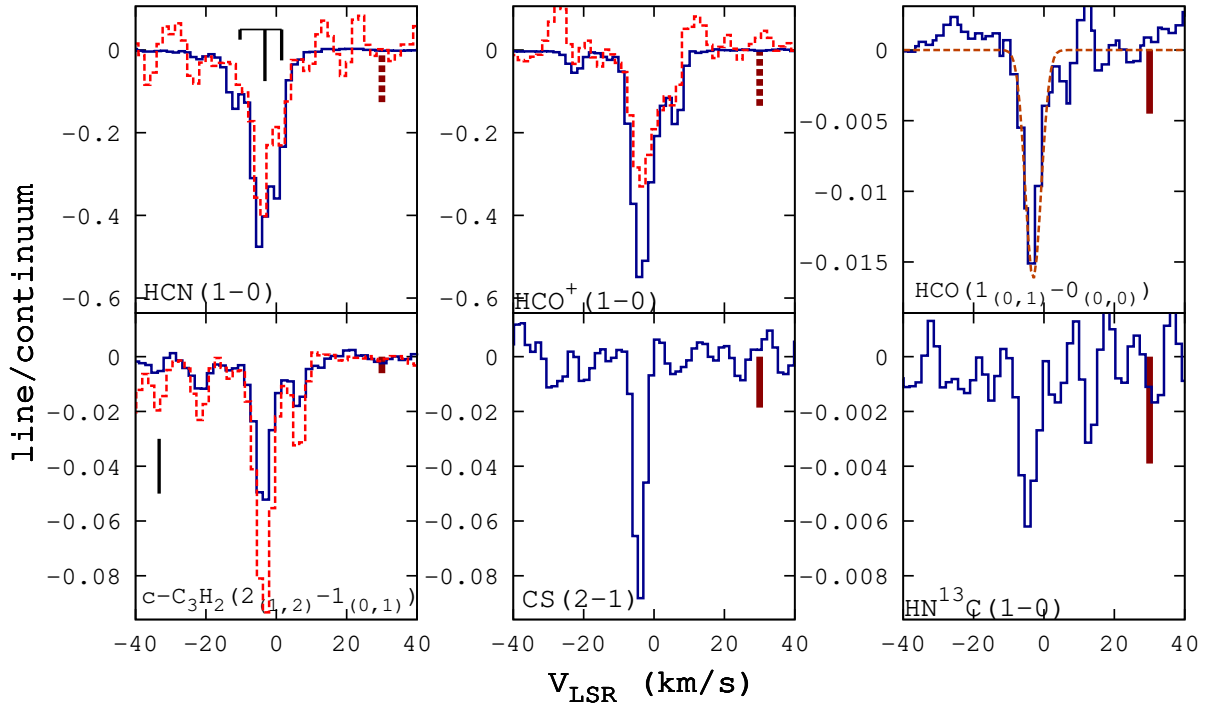} 
 \end{center}
\caption{Spectra of all molecules detected toward J1717$-$337. 
  The red dotted lines superimposed on the blue solid spectra of HCN(1--0) and HCO$^+$(1--0) 
  are the H$^{13}$CN(1--0) and H$^{13}$CO$^+$(1--0) spectra, respectively, 
  both of which are multiplied by a factor of 30
  (the $3\sigma$ noise levels indicated by the red dotted bars 
  are those of the $^{13}$C isotopologues, 
  which are also multiplied by 30; 
  those of the $^{12}$C isotopologues are too small to be shown). 
  The intrinsic line strength ratios of the HCN(1--0) hyperfine components 
  are also shown beside the former. 
  An example of a Gaussian fit result, 
  for the HCO(1$_{(0,1)}$--0$_{(0,0)}$) ($J = 3/2-1/2, \ F = 2-1$) absorption line, 
  is shown in an orange dotted line superimposed on the blue solid spectrum.
  The C$_2$H(1--0) ($J = $ 3/2--1/2, $F =$ 2--1) spectrum is represented by the red dotted line, 
  superimposed on the c-C$_3$H$_2$(2$_{(1,2)}$--1$_{(0,1)}$) blue solid spectrum.
  The black bar on the left side indicates the HCS$^+$(2--1) line frequency, 
  although the HCS$^+$(2--1) detection is marginal and should be treated with some caution.
  }\label{fig2}
\end{figure*}

\begin{figure*}[t]
 \begin{center}
  \includegraphics[width=160mm]{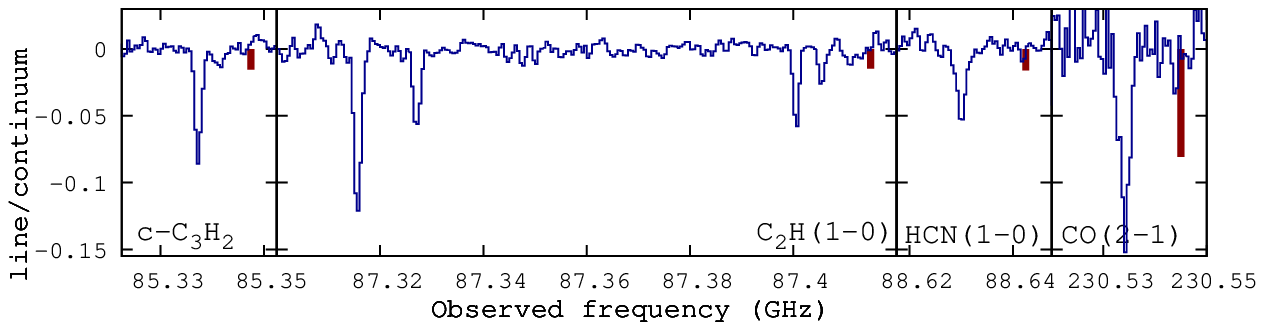} 
 \end{center}
\caption{Spectra of all molecules detected toward J1625$-$254: 
c-C$_3$H$_2$(2$_{(1,2)}$--1$_{(0,1)}$), C$_2$H(1--0), HCN(1--0), and CO(2--1), from left to right.}
\label{fig3}
\end{figure*}

The optical depth and column density of each molecule shown in table \ref{tab3} are calculated 
using equations (1) and (2) of Greaves and Nyman (\yearcite{GN96})
(see Appendix 2 for details), 
assuming that the excitation temperatures ($T_{\mathrm{ex}}$) 
of the detected molecules are 2.73 K.
This means that these molecules are taken to be in equilibrium with the CMB, 
as has been widely assumed in the previous studies on Galactic molecular absorption systems 
(e.g., \cite{GN96}; \cite{LL98}; \cite{Liszt+14}).
Further, this has been confirmed for some diffuse media 
observed toward Galactic star-forming regions (\cite{Godard+10}).
With this assumption, we can derive optical depths
from the line/continuum flux density ratios, 
which are free from flux calibration uncertainty ($\sim 10$\%).
However, this assumption should be treated with caution 
if the ISM in question is located in a photodissociation region (PDR, see section 4).
Even if $T_{\mathrm{ex}}$ is higher than 2.73 K, the measured column density ratios, 
such as those of $N$(HCO)/$N$(H$^{13}$CO$^+$) discussed in section 4, 
do not change dramatically for molecules with similar $T_{\mathrm{ex}}$.
For example, if $T_{\mathrm{ex}} = 10$ K, 
the inferred column densities of all species increase by almost the same factor ($\sim 8$)
for transition from the ground states; 
this does not change the abundance ratios no more than 10\%.
Such a discrepancy is negligible compared with the errors associated with the flux calibration, 
root-mean-square (RMS) noise, and line fits.
Further analysis of the impact of $T_{\mathrm{ex}}$ and the uncertainty regarding column densities 
and their ratios will be conducted in a forthcoming paper (Ando et al., in preparation). 
In addition, the frequency ranges we analyze include 
the line frequencies of various well-known molecules that are not detected in this work.
Table \ref{tab4} summarizes the $3\sigma$ upper limits 
of the peak optical depths and column densities of relatively major undetected species.

\section{Results}
\begin{figure}[t]
 \begin{center}
  \includegraphics[width=8cm]{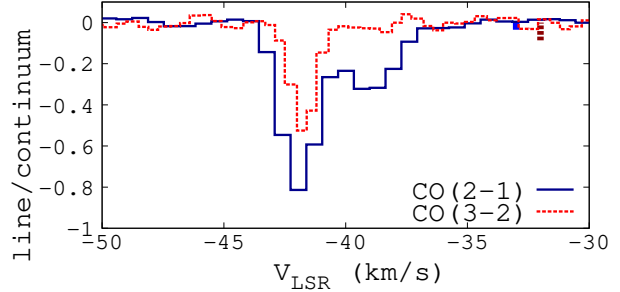} 
 \end{center}
\caption{J1604$-$446 spectra. The blue solid and red dotted lines represent 
the CO(2--1) and CO(3--2) spectra, respectively.}\label{fig4}
\end{figure}

Of the 36 calibrators analyzed in this work,
molecular absorption lines are detected toward the four objects listed in table $\ref{tab2}$.
All of these are found to be Galactic in origin, based on consideration of the line velocities.
While the absorption line system found toward NRAO530 
has been identified in previous studies (e.g., \cite{LL96}, \cite{LL98}), 
the other three systems, toward J1717$-$337, J1625$-$254, and J1604$-$446
are newly detected in this work.
The spectra of the four sources are shown in figures \ref{fig1}-\ref{fig4},
and each object is described in detail below.

\textbf{J1717$-$337} is known to exhibit Galactic HI absorption (\cite{Dickey+83}), 
although molecular absorption has never been reported.
We detect the ten species shown in table $\ref{tab2}$, 
including relatively rare species such as HN$^{13}$C and HCS$^+$, 
all of which are new detections in this system.
In addition, four HCO(1$_{(0,1)}$--1$_{(0,0)}$) and six C$_2$H(1--0) hyperfine structure lines
are clearly detected.
Four velocity components are identified distinctly, as shown in figures \ref{fig1} and \ref{fig2}.
Further discussions of the nature of each molecule detected toward J1717$-$337
and of the $^{12}$C/$^{13}$C isotope ratio will be presented in a forthcoming paper
(Ando et al., in preparation).

\textbf{J1625$-$254} has not been investigated in terms of Galactic absorption,
except for a previous search for high-$z$ molecular absorption (\cite{WC96}), 
with non-detection being reported in terms of this line-of-sight.
Hence, all four species reported in this work are new detections.
Further, HCO and H$^{13}$CO$^+$ are not detected,  
which is significantly different to the profile of J1717$-$337.
Only one velocity component is identified, as shown in figure \ref{fig3}.

\textbf{J1604$-$446} has never been referred to in previous studies.
Thus, this is a new Galactic molecular absorption system. 
As shown in figure \ref{fig4}, absorption lines indicating two CO transitions are detected, 
with the weaker line corresponding to CS(2--1). 
Two velocity components are identified in the CO(2--1) spectrum, 
while only the $V_{\mathrm{LSR}} = -42$ km/s component is confirmed in that of CO(3--2).
Using a rotation diagram with the two transition lines, 
the $T_{\mathrm{ex}}$ of the main CO component is calculated to be 6.2 K.
However, this analysis assumes local thermodynamic equilibrium (LTE) and an optically-thin limit, 
which should be treated carefully in the case of CO lines with moderate optical depths.

\textbf{NRAO\,530} is an extensively-studied Galactic molecular absorption system 
(e.g., \cite{LL96}; \cite{LL98}), 
but its HCO absorption lines are newly detected (see section 4).
A single velocity component is identified, the velocity of which ($V_{\mathrm{LSR}} = 6$ km/s)
is consistent with the values reported in previous studies (\cite{LL96}; \cite{LL00}).

\section{Discussion}
\begin{figure*}[t]
 \begin{center}
  \includegraphics[width=16cm]{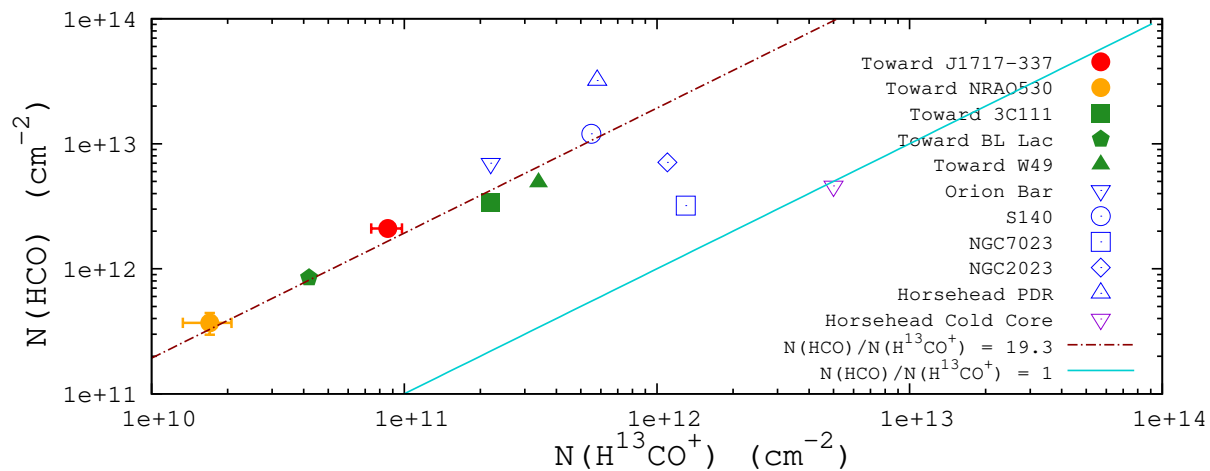} 
 \end{center}
\caption{Column densities of HCO vs. H$^{13}$CO$^+$.
The column densities measured toward J1717-337 and NRAO530 in this work 
are plotted in red and orange circles, respectively, with error bars. 
The data toward 3C111, BL Lac, and W49 (40 km s$^{-1}$ feature) are from \citet{Liszt+14}, 
and are plotted using green symbols. 
The data for four well-known PDRs (Orion Bar, S140, NGC7023 and NGC2023), 
which are represented by blue symbols, are from \citet{Schilke+01}, 
although $N$(H$^{13}$CO$^+$) are derived from $N$(H$^{12}$CO$^+$), 
assuming $N$(H$^{12}$CO$^+$)/$N$(H$^{13}$CO$^+$) $= 77$ (\cite{Schilke+01}). 
The data for the Horsehead PDR (blue) and Horsehead Cold Core (purple) 
correspond to the data for the HCO and DCO$^+$ peaks of the Horsehead Nebula, respectively, 
which were obtained by \citet{Gerin+09}. 
As a whole, the filled symbols represent absorption observation results, 
while open symbols are emission studies data.
The red dotted and cyan solid lines correspond to the values of HCO to H$^{13}$CO$^+$ 
column density ratios of 19.3 (the average of the ratios measured in diffuse ISM 
toward the five continuum sources where HCO is detected) and 1 
(the threshold of the PDR-like environment in the ISM, \cite{Gerin+09}), respectively.}\label{fig5}
\end{figure*}

Among the various species detected in this study, HCO is the most noteworthy.
Although its emission has been investigated in HII region interfaces (\cite{Schenewerk+88})
and in the nearby starburst galaxies, M82 (\cite{GB+02}) and NGC253 (\cite{Martin+09}),
HCO absorption in diffuse media has only been reported 
along the lines-of-sight toward two bright quasars (BL Lac and 3C111) 
and toward the W49 Galactic HII region (\cite{Liszt+14}).
Therefore, the HCO absorptions toward J1717$-$337 and NRAO530 
are the fourth and fifth detections, 
and this work almost doubles the number of HCO absorption samples in diffuse media.

The detection of HCO absorption itself is quite rare, but it has certain implications.
The HCO in diffuse ISM can be used as a PDR tracer 
(\cite{Schenewerk+88}; \cite{Gerin+09}; \cite{Martin+09}), 
because of the diffuse media environment.
Although there is a moderate density ($n_{\mathrm{H}_2} \sim 10^1 - 10^2$ cm$^{-3}$)
of molecular hydrogen (H$_2$) gas in a diffuse medium (\cite{Draine11}), 
such a medium is expected to be illuminated by far-UV (FUV) radiation from distant OB stars
(\cite{HT99}).
In such conditions, carbon ionized by FUV radiation should be present, 
as it has been suggested that ionized carbon (C$^+$) is 
the main carbon reservoir in neutral diffuse ISM (\cite{Gerin+15}).
In a diffuse medium, therefore, it is probable that HCO is formed in the gas phase, 
through the reaction
\begin{equation}
\mathrm{O + CH_2 \rightarrow HCO + H}
\end{equation}
(\cite{Gerin+09}).
CH$_2$ is formed from C$^+$ and H$_2$ through several reactions (\cite{Schenewerk+88}), 
and such chemistry is realized in the environment where C$^+$ and H$_2$ coexist.
Such an environment is comparable to that of classical PDRs, 
i.e., interface regions between HII regions and molecular clouds (\cite{HT99}).
HII regions are directly illuminated by UV radiation, 
and molecular clouds are dense neutral regions shielded from such radiation.

Furthermore, the $N$(HCO)/$N$(H$^{13}$CO$^+$) column density ratio can be used 
as a diagnostic of the presence of an FUV radiation field, 
and its high ratio ($\gtrsim 1$) suggests 
the presence of ongoing FUV photochemistry (\cite{Gerin+09}).
Although H$^{13}$CO$^+$ is known to be sensitive to high-density gas 
($n_{\mathrm{H}_2} \gtrsim 10^5$ cm$^{-3}$; \cite{Onishi+02}; \cite{Gerin+09}),
taking the H$^{13}$CO$^+$ to HCO column density ratio 
is rather useful in accurately estimating the relative abundance of HCO, 
because their close proximity in frequency 
(H$^{13}$CO$^+$ and HCO at 86.671 and 86.754 GHz)
allows us to observe them simultaneously (e.g., \cite{GB+02}; \cite{Liszt+14}).
Figure \ref{fig5} shows the HCO and H$^{13}$CO$^+$ column densities toward the two calibrators, 
compared with the values measured on the lines-of-sight of the sources 
toward which HCO absorption was previously detected (\cite{Liszt+14}), 
along with the values in several PDRs (\cite{Schilke+01}; \cite{Gerin+09}).
The $N$(HCO)/$N$(H$^{13}$CO$^+$) ratios recorded toward J1717$-$337 and NRAO530 
are $24 \pm 4$ and $23 \pm 7$, respectively, 
which are comparable to the values for well-known PDRs (\cite{Schilke+01});
this suggests that the media observed toward these sources are in PDR-like environments.

In terms of the heating mechanisms of diffuse ISM, 
a molecular absorption study has suggested that the gas-phase chemistry of 
some diffuse media is driven by turbulent dissipation (\cite{Godard+10}).
Meanwhile, observations of the [CII] fine structure line (\cite{Ingalls+02})
and pure-rotational transitions of molecular hydrogen (\cite{Ingalls+11}) have demonstrated
the role of FUV photons in diffuse media, 
as it is expected that UV radiation from distant OB stars 
can penetrate and illuminate diffuse ISM (\cite{HT99}).
However, there are a limited number of studies that constitutes observational evidences
that FUV photons drive PDR-like chemistry in diffuse molecular gas.
This is primarily because, in diffuse ISM, minor molecular species including PDR-tracers
cause only minimal emission as a result of their low densities and low excitation states.
Therefore, it is remarkable that HCO enhancement comparable to that of PDRs,
which suggests a PDR-like environment, is found in diffuse ISM. 
Importantly, this result observationally illustrates 
the role of FUV photons in triggering the PDR-like chemistry, 
and is expected to help us to further understand the heating mechanisms of such media.
More detailed investigation into the chemistry and physics of PDR-like diffuse ISM
will be presented in a forthcoming paper (Ando et al., in preparation), 
along with the data on the molecular absorption of higher transition lines, 
the observation of which is planned in ALMA Cycle 3.

As three Galactic molecular absorption systems are newly detected in this study, 
the use of ALMA calibrators as potential sources in the quest for molecular absorption systems
is proposed.
Moreover, the detection of HCO absorption also proves the power of ALMA:
just a few minutes' integration with ALMA almost doubles 
the number of HCO absorption samples in diffuse media, 
and provides observational support 
for the PDR-like chemistry of diffuse ISM processed by FUV radiation.
ALMA follow-up observations of the molecular absorption systems that are likely to be discovered 
through ALMA-calibrator-based searches 
will constitute not only an efficient manner of investigating the chemical state of Galactic diffuse ISM, 
but also the best shortcut to extensive comprehension of the detailed nature of ISM as a whole, 
from our Galaxy to the high-$z$ universe.

\begin{ack}
We thank the anonymous referee for insightful comments, 
which helped us to improve the paper.
This paper makes use of the following ALMA data: ADS/JAO.ALMA \#2011.0.00217.S., 
2011.0.00259.S., 2011.0.00351.S., 2011.0.00524.S., 2011.0.00531.S., and 2011.0.00733.S.
ALMA is a partnership of ESO (representing its member states), NSF (USA) and NINS (Japan), 
together with NRC (Canada), NSC, ASIAA (Taiwan) and KASI (Republic of Korea), 
in cooperation with the Republic of Chile. 
The Joint ALMA Observatory is operated by ESO, NAOJ and NRAO.
\end{ack}

\appendix
\section{List of targets analyzed in this work}
Table \ref{tab5} lists the 36 calibrator sources analyzed in this work, 
and shows the observation projects that contain these sources 
and the bands and frequency ranges analyzed in this study, 
and indicates whether or not absorption lines are detected.

\section{Derivation of molecule optical depths and column densities}
The peak optical depth of each molecule shown in table \ref{tab3} is derived from 
\begin{equation}
\tau_\nu = -\ln{\left[1 - \frac{T_{\mathrm{MB}}}{J(T_{\mathrm{ex}}) - J(T_{\mathrm{CMB}}) 
- T_{\mathrm{cont}}} \right]},
\end{equation}
where $T_{\mathrm{MB}}$ is the main beam brightness temperature of the line, 
$T_{\mathrm{CMB}}$ is the CMB temperature (2.73 K), 
$T_{\mathrm{ex}}$ is the excitation temperature of the line, 
$T_{\mathrm{cont}}$ is the continuum antenna temperature, and 
\begin{equation}
J(T) = \frac{h\nu}{k} \frac{1}{\exp(\frac{h\nu}{kT}) - 1}.
\end{equation}
Equation (A1) is derived from equation (1) of Greaves and Nyman (\yearcite{GN96}).
Note that $T_{\mathrm{MB}}$ is negative for an absorption line, 
as it is defined as the main beam brightness temperature of the line 
measured in the continuum-subtracted spectrum.
As was commonly assumed in the previous studies on Galactic molecular absorption systems
(e.g., \cite{GN96}; \cite{LL96}; \cite{Liszt+14}), 
we assume that the $T_{\mathrm{ex}}$ of the detected molecules is 2.73 K,
which has the advantage that $\tau_\nu$ can be calculated 
from the line/continuum flux density ratio in a straightforward manner, 
because $J(T_{\mathrm{ex}})$ in equation (A1) is offset by $J(T_{\mathrm{CMB}})$.

The total column density of each molecule is calculated from
\begin{eqnarray}
N_{\mathrm{total}} &=& \frac{3h}{8 \pi^3 S \mu^2} 
\frac{\displaystyle Q(T_{\mathrm{ex}}) \exp{\left(\frac{E_l}{k T_{\mathrm{ex}}} \right)}}
{\displaystyle \left[1 - \exp{\left(-\frac{h\nu}{k T_{\mathrm{ex}}}\right)} \right]} \int \tau dv 
\nonumber \\
&\equiv& F(T_{\mathrm{ex}}) \int \tau dv,
\end{eqnarray}
where $S$ is the intrinsic line strength, 
$\mu$ is the permanent electric dipole moment of the molecule, 
$Q(T)$ is the partition function of temperature $T$, 
and $E_l$ is the lower energy level of the transition (\cite{GN96}).
For each molecule, $F(T_{\mathrm{ex}})$, 
which is the conversion factor used to derive the column densities from the integrated optical depths,
is shown in table \ref{tab6}, assuming a $T_{\mathrm{ex}}$ of 2.73 K.
As the profile of each line is assumed to be a Gaussian 
of peak optical depth $\tau_\nu$ and FWHM $\Delta V$, 
the integrated optical depths are calculated as
\begin{equation}
\int \tau dv = \sqrt{\frac{\pi}{4\ln{2}}} \tau_\nu \Delta V
\end{equation}
(\cite{Godard+10}). Therefore the column density of a certain molecule can be derived 
from the measured $\tau_\nu$ and $\Delta V$ of its absorption line, 
if $T_{\mathrm{ex}}$ is assumed.

\begin{table*}[p]
  \tbl{List of all targets analyzed in this work.}{
    \begin{tabular}{lcclc}
   ~~ \\
\hline\noalign{\vskip3pt} 
       \hspace{0.5cm} Object name & Project & Band 
       & \hspace{2.4cm} Frequency range (GHz) & Detection  \\
       [2pt] \hline\noalign{\vskip3pt}
       J0116-116 & 2011.0.00099.S & 3 & 112.58-114.46 & N\\
       J0132-169 & 2011.0.00061.S & 3 & 96.38-100.25, 108.42-111.29 & N\\
       J0137-245 & 2011.0.00172.S & 3 & 85.70-87.57, 87.63-89.40, 97.58-101.32& N \\
       J0217+017 & 2011.0.00083.S & 6 & 341.01-344.70, 352.89-356.64& N \\
       & 2011.0.00243.S & 7 & 240.25-241.95, 242.61-244.20, 254.65-256.31, 257.09-258.73& N \\
       J0334-401 & 2011.0.00108.S & 3 & 85.40-89.10, 97.27-100.92 & N \\
       J0339-017 & 2011.0.00061.S & 3 & 96.07-97.94, 98.07-99.94, 108.11-110.94 & N \\
       J0403-3605 & 2011.0.00099.S & 3 & 110.36-112.24 & N \\
       J0423-013 & 2011.0.00061.S & 3 & 96.07-97.94, 98.07-99.94, 108.11-110.94 & N \\
       J0455-462 & 2011.0.00208.S & 7 & 
       352.51-354.38, 354.47-356.34, 340.51-342.38, 342.47-344.34 & N \\
       J0522-364 & 2011.0.00108.S & 3 & 85.40-89.10, 97.27-100.92& N \\
       & 2011.0.00099.S & 3 & 108.66-110.53& N \\
       & 2011.0.00099.S & 3 & 110.83-112.71& N \\
       & 2011.0.00108.S & 7 & 340.98-344.44, 352.60-356.19& N \\
       J0538-4405 & 2011.0.00170.S & 6 & 233.63-235.51 & N \\
       J0607-085 & 2011.0.00170.S & 6 & 216.25-218.84, 233.63-236.02 & N \\
       & 2011.0.00170.S & 6 & 245.32-247.20 & N \\
       J0637-752 & 2011.0.00273.S & 3 & 100.09-101.96, 102.05-103.92, 112.21-115.92 & N \\
       & 2011.0.00471.S & 3 & 85.94-86.88, 88.36-89.29, 97.43-98.36, 98.47-99.40& N \\
       J0854+201& 2011.0.00307.S & 7 & 352.04-353.92 & N \\
       J0909+013 & 2011.0.00307.S & 7 & 352.04-353.92 & N \\
       J1038-5311 & 2011.0.00497.S & 3 & 84.77-88.34 & N \\
       J1107-448 & 2011.0.00497.S & 3 & 84.77-88.34 & N \\
       J1130-148 & 2011.0.00497.S & 3 & 86.47-88.34 & N \\
       & 2011.0.00020.S & 7 & 300.01-303.65, 311.94-315.77 & N \\
       & 2011.0.00525.S & 7 & 353.79-355.66 & N \\
       J1256-057 (3C279) & 2011.0.00351.S & 3 & 84.42-88.06, 96.68-100.26 & N \\
       & 2011.0.00121.S & 7 & 319.69-321.56 & N \\
       J1325-430 & 2011.0.00121.S & 7 & 319.69-321.56 & N \\
       J1329-5608 & 2011.0.00121.S & 7 & 319.84-321.71 & N \\
       J142739-330612 & 2011.0.00099.S & 3 & 109.68-111.55 & N \\
       J1427-421 & 2011.0.00524.S & 3 & 345.60-346.10 & N \\
       J1517-243 & 2011.0.00121.S & 7 & 316.59-318.47, 320.48-322.35 & N \\
       J1540+147 & 2011.0.00175.S & 7 & 334.09-335.81, 335.97-337.69, 346.09-349.52& N \\
       J1604-446 & 2011.0.00733.S & 3 & 95.06-98.83, 107.06-108.94, 109.06-110.94 & Y \\
       & 2011.0.00733.S & 6 & 229.61-231.48, 231.60-233.48 & Y \\
       & 2011.0.00524.S & 7 & 344.94-348.23, 356.41-359.62 & Y \\
       J1625-254 & 2011.0.00259.S & 3 & 85.16-87.03, 87.05-88.93, 97.16-99.03, 99.05-100.93 & Y \\
       & 2011.0.00531.S & 6 & 219.48-219.61, 220.34-220.44, 230.46-230.56, 232.35-232.48 & Y \\
       & 2011.0.00259.S & 7 & 330.17-332.04, 332.07-333.94, 342.17-344.04, 344.07-345.94 & N \\
       J1700-261 & 2011.0.00017.S & 3 & 85.90-87.77 & N \\
       J1717-337 & 2011.0.00217.S & 3 & 
       86.26-88.14, 88.15-90.03, 98.19-100.07, 100.15-102.03 & Y \\
       & 2011.0.00351.S & 3 & 84.42-88.06, 96.68-100.26 & Y \\
       J1733-130 (NRAO530) & 2011.0.00217.S & 3 & 
       86.26-88.14, 88.15-90.03, 98.19-100.07, 100.15-102.03 & Y \\
       & 2011.0.00733.S & 3 & 95.06-96.94, 96.95-98.83, 107.06-108.94, 109.06-110.94 & Y \\
       & 2011.0.00733.S & 6 & 229.61-231.48, 231.60-233.48 & Y \\
       J1751+096 & 2011.0.00526.S & 7 & 334.71-336.59 & N \\
       J1924-292 & 2011.0.00108.S & 3 & 85.40-89.10, 97.27-100.92 & N \\
       & 2011.0.00351.S & 3 & 84.42-88.06, 96.68-100.26 & N \\
       & 2011.0.00208.S & 7 & 340.51-342.38 & N \\
       & 2011.0.00526.S & 7 & 334.76-336.63 & N \\
       J1945-552 & 2011.0.00046.S & 3 & 99.64-103.35, 111.52-115.23& N \\
       J2253+161 (3C454.3) & 2011.0.00083.S & 7 & 341.01-344.70, 352.89-356.64& N \\
       J2258-279 & 2011.0.00061.S & 3 & 96.38-98.25, 98.37-100.25, 108.42-111.29 & N \\
       J2333-237 & 2011.0.00172.S & 3 & 85.70-87.57, 87.63-89.40, 97.58-101.32 & N \\ [2pt] 
      \hline
    \end{tabular}}\label{tab5}
\end{table*}

\begin{table*}[t]
  \tbl{Conversion factors used to derive column densities from integrated optical depths,
  assuming that $T_{\mathrm{ex}} = 2.73$ K. 
  The intrinsic line strengths $S$ and permanent electric dipole moments $\mu$ of each molecule, 
  which are used to derive the conversion factors, 
  are taken from \citet{Muller+11} and \citet{Muller+14},
  with original reference to data from
  The Cologne Database for Molecular Spectroscopy\footnotemark[$*1$] (CDMS) 
  and the Jet Propulsion Laboratory Molecular Spectroscopy\footnotemark[$*2$] (JPL).}
  {\begin{tabular}{cccccl}
   ~~ \\
\hline\noalign{\vskip3pt} 
       \multirow{2}{*}{Species} & Frequency & Intrinsic line strength 
       & Electric dipole moment & $F(T_{\mathrm{ex}} = 2.73 \ \mathrm{K})$ 
       & \multirow{2}{*}{\hspace {1.3cm} Notes} \\
       & (GHz) & $S$ & $\mu$ (debye) & (cm$^{-2}$ km$^{-1}$ s) & \\
       [2pt] \hline\noalign{\vskip3pt}
      HC$^{18}$O$^+$(1--0) & 85.1622 & 1.00 & 3.90 
      & $1.17 \times 10^{12}$ & For undetected lines\\
      c-C$_3$H$_2$(2$_{(1,2)}$--1$_{(0,1)}$) & 85.3389 & 4.50 & 3.43 
      & $4.64 \times 10^{12}$ & \\
      HCS$^+$(2--1) & 85.3479 & 2.00 & 1.96 & $8.62 \times 10^{12}$ & \\
      HC$^{15}$N(1--0) & 86.0550 & 1.00 & 2.99 & $1.98 \times 10^{12}$ & For undetected lines \\
      H$^{13}$CN(1--0) & 86.3402 & 1.00 & 2.99 & $1.97 \times 10^{12}$ & $F = 2-1$\\
      HCO(1$_{(0,1)}$-0$_{(0,0)}$) & 86.6708 & 0.42 & 1.36 & 
      $2.26 \times 10^{13}$ & $J = 3/2-1/2, \ F = 2-1$\\ 
      H$^{13}$CO$^+$(1--0) & 86.7543 & 1.00 & 3.90 & $1.15 \times 10^{12}$ & \\ 
      SiO(2--1) & 86.8470 & 2.00 & 3.10 & $3.43 \times 10^{12}$ & For undetected lines \\ 
      HC$^{17}$O$^+$(1--0) & 87.0575 & 1.00 & 3.90 & 
      $1.14 \times 10^{12}$ & For undetected lines \\
      HN$^{13}$C(1--0) & 87.0909 & 1.00 & 3.05 & $1.87 \times 10^{12}$ & \\
      C$_2$H(1--0) & 87.3169 & 0.42 & 0.80 & $6.45 \times 10^{13}$ 
      & $J = 3/2-1/2, \ F = 2-1$ \\
      HCN(1--0) & 88.6318 & 1.00 & 2.99 & $1.91 \times 10^{12}$ & $F = 2-1$ \\
      HCO$^+$(1--0) & 89.1885 & 1.00 & 3.90 & $1.11 \times 10^{12}$ & \\
      HOC$^+$(1--0) & 89.4874 & 1.00 & 2.77 & $2.20 \times 10^{12}$ & For undetected lines \\
      C$^{34}$S(2--1) & 96.4130 & 2.00 & 1.96 & $8.25 \times 10^{12}$ & For undetected lines \\
      CS(2--1) & 97.9810 & 2.00 & 1.96 & $8.09 \times 10^{12}$ & \\
      C$^{18}$O(2--1) & 219.5604 & 2.00 & 0.11 & $3.69 \times 10^{15}$ & For undetected lines \\
      $^{13}$CO(2--1) & 220.3987 & 2.00 & 0.11 & $3.68 \times 10^{15}$ & For undetected lines \\
      CO(2--1) & 230.5380 & 2.00 & 0.11 & $3.59 \times 10^{15}$ & \\
      CO(3--2) & 345.7960 & 3.00 & 0.11 & $1.36 \times 10^{17}$ & \\ [2pt]
      \hline
    \end{tabular}}\label{tab6}
\begin{tabnote}
    \hangindent6pt\noindent
\hangindent0pt\noindent
\hbox to8pt{\footnotemark[$*1$]\hss}\unskip
\verb|https://www.astro.uni-koeln.de/cdms/|
\newline
\hangindent0pt\noindent
\hbox to8pt{\footnotemark[$*2$]\hss}\unskip
\verb|http://spec.jpl.nasa.gov/ftp/pub/catalog/catdir.html|
\end{tabnote}
\end{table*}

\end{document}